\documentstyle[aps,multicol,epsfig]{revtex}
\begin{document}
\draft

\title{Dopant-induced crossover from 1D to 3D charge transport in conjugated
polymers}

\author{J.A. Reedijk, H.C.F. Martens, H.B. Brom}

\address{Kamerlingh Onnes Laboratory, Leiden University, P.O.Box
9504, 2300 RA Leiden, The Netherlands}

\author{M.A.J. Michels}

\address{Department of Applied Physics, Eindhoven University of
Technology, and Dutch Polymer Institute, PO Box 513, 5600 MB Eindhoven,
The Netherlands}

\date{April 29, 1999, Phys. Rev. Letters, to appear November 1999}

\maketitle

\begin{abstract}

The interplay between inter- and intra-chain charge transport in bulk
polythiophene in the hopping regime has been clarified by studying
the conductivity $\sigma$ as a function of frequency $\omega/2\pi$ (up to
3~THz), temperature $T$ and doping level $c$. We present a model which
quantitatively explains the observed crossover from quasi-one-dimensional
transport to three-dimensional hopping conduction with increasing doping
level. At high frequencies the conductivity is dominated by charge transport
on one-dimensional conducting chains.


\end{abstract}

\begin{multicols}{2}

\settowidth{\columnwidth}{aaaaaaaaaaaaaaaaaaaaaaaaaaaaaaaaaaaaaaaaaaaaaaaaa}

The charge transport mechanisms in conjugated polymers, although extensively
studied over the last two decades, are still far from completely understood.
Not only the behavior around the insulator-to-metal transition (IMT), which
can be induced in several polymer materials upon appropriate doping, but also
the nature of hopping transport in the deeply insulating regime are not yet
resolved. While some studies indicate that transport is dominated by hops
between three-dimensional (3D), well conducting
regions\cite{Pelster94,Zuppiroli94}, in other cases the strongly
one-dimensional (1D) character of the polymer systems appears to be a crucial
factor\cite{Wang90,Wang91,Li93}.

In investigating the nature of hopping transport in conjugated polymers,
studying the temperature and doping level dependence of the DC conductivity is
an important tool. Since the DC conductivity is determined by the weakest
links in the conducting path spanning the sample, the study of
$\sigma^{}_{DC}(T)$ gives insight in the slowest relevant transport processes
in the system.

On the insulating side of the IMT, the DC conductivity is predicted by many
models to follow the well-known hopping expression

\begin{equation}
\sigma^{}_{DC} = \sigma^{}_0 e^{-(T^{}_0/T)^\gamma}
\label{mott}
\end{equation}

\noindent{where the value of $\gamma$ and the interpretation of $T^{}_0$
depend on the details of the model. The original Mott theory for 3D variable
range hopping with a constant density of states (DOS) at the Fermi energy
predicts $\gamma=1/4$\cite{Mott69}, while several modifications of the model
have been proposed to describe the frequently observed value $\gamma=1/2$.
Studying the dependence of $\gamma$ and $T^{}_0$ on doping level $c$ provides
the opportunity to discriminate between the various hopping models and extract
parameters determining the conductive properties like the DOS and the
localization length.}

While the DC conductivity is sensitive to the slowest transport processes,
the AC conductivity $\sigma(\omega)$ provides information about
processes occurring at time scales $\tau \approx \omega^{-1}$.
Especially in conjugated polymers, where intra-chain and inter-chain
transition rates may differ by orders of magnitude, knowledge of
$\sigma(\omega)$ at high frequencies can help to clarify the properties of
charge transport on a polymer chain.

In this Letter, we present a systematic study of the charge transport in a
conjugated polymer far away from the IMT, as a function of frequency,
temperature and doping level. By selecting a polymer system with very low
inter-chain mobility, a separation of inter-chain and intra-chain
contributions to the conductivity can be made when the applied frequency is
varied over 12 decades. At low frequencies, transport between chains is
studied, while at frequencies well above the inter-chain transition rate,
intra-chain conduction is probed.

{\sl Experimental --}
The experiments were performed on discs (thickness between 0.4 and 1.0 mm)
of pressed powders of the conjugated polymer
poly(3,4-di-[(R,S)-2-methyl\-butoxy]thiophene), abbreviated as PMBTh, the
synthesis of which is described elsewhere\cite{Voss96}. The samples were doped
with FeCl$_3$ in a dichloromethane solution at doping levels $0.01<c<0.22$;
here $c$ is the number of doped carriers per thiophene ring, which was
determined with the aid of M\"ossbauer spectroscopy\cite{Goossens,TUE}. After
doping, the solvent was evaporated and the resulting powders were vacuum dried
overnight.  The conductivity of the samples remained unchanged in an ambient
atmosphere for several weeks. Contact resistances were less than 5\% of the
sample resistance\cite{film}.  The conductivity data were taken in the range
5~Hz--3~THz with the aid of several experimental methods, which are described
elsewhere\cite{Martens99}.  The conductivity in the range 0.3--3~THz was
determined from the transmission measured with a Bruker FTIR spectrometer.

{\sl DC conductivity --}
The temperature-dependent static conductivity of samples with doping levels
ranging from 0.01--0.22 is plotted in Fig.~\ref{DCconductivity}. Here
$\log\sigma^{}_{DC}$ is plotted vs. $T^{-1/2}$, so that data sets following
Eq.~(\ref{mott}) with $\gamma =1/2$ fall on a straight line. In the inset, the
value of $\gamma$ is plotted vs.  $c$. The exponent $\gamma$ has
been determined by plotting the logarithm of the reduced activation energy
$W = d\ln\sigma/d\ln T$ versus $\ln T$ and fitting it to a straight line; the
slope of this line gives $\gamma$\cite{Zabrodskii84}. The data show a clear
transition in the DC behavior as a function of doping level around $c^{}_0 =
0.12$. For low doping $c < c^{}_0$, $\gamma$ values are grouped around $1/2$,
whereas for $c > c^{}_0$, $\gamma$ is close to the Mott value $1/4$. The
conductivity data for $c < c^{}_0$ are now fitted to Eq.~(\ref{mott}) with
$\gamma$ fixed at $1/2$ (solid lines), while the data for $c > c^{} _0$ are
fitted with a fixed $\gamma=1/4$ (dashed lines). Note that as the conductivity
is many of orders of magnitude below the minimum metallic conductivity $\sim$
100 S/cm and depends very strongly on $T$, PMBTh is far away from the IMT at
all doping levels.

Many authors\cite{Zuppiroli94,Wang90,Wang91,Li93} have reported the
conductivity in conjugated polymers to follow Eq.~(\ref{mott}) with
$\gamma=1/2$, and several models have been proposed to explain this value.
In disordered systems, the single particle DOS around the
Fermi energy has a parabolic shape when long-range Coulomb interactions
between charge carriers are dominant\cite{Shklovskii85}. Inserting a quadratic
DOS in the original Mott argument yields the exponent $\gamma=1/2$. Data on
various conjugated polymers close to the IMT have been interpreted within this
Efros and Shklovskii Coulomb-gap model\cite{Yoon95}.  Alternatively, it has
been argued that polymer materials can be viewed as 3D
granular metallic systems when the strong, inhomogeneous disorder in polymer
materials leads to the formation of well conducting 3D regions separated by
poorly conducting barriers\cite{Pelster94,Zuppiroli94}. For granular metals,
$\sigma^{}_{DC}(T)$, though still not completely understood, is widely
accepted to follow Eq.~(\ref{mott}) with $\gamma=1/2$\cite{Abeles75}.
Close to the IMT both models predict a crossover from $\gamma=1/2$ to
$\gamma=1/4$. Such transitions have indeed been observed in a number of
systems \cite{Yoon95,Finlayson87,Castner91}. However, because PMBTh is far
away from the IMT at all $c$, these models cannot be applied here.

To explain our data we will use another approach, which is an extension of the
quasi-1D hopping model of Nakhemdov {\sl et al.}\cite{Nakhmedov89}. In their
model, the charge carriers are supposed to be strongly localized on single
chains (or 1D bundles of chains). Variable-range hopping is possible along the
chains, while perpendicular to the chains with vanishingly small inter-chain
overlap only nearest-neighbor hopping is allowed. Although the quasi-1D model
predicts $\gamma=1/2$ strictly speaking only for the anisotropic effective
conductivity perpendicular to the chains, it has been successfully applied to
randomly oriented polymer systems\cite{Wang90}. In a closely related approach,
the polymer system is viewed as a fractal structure with a fractal dimension
$d_f$ slightly greater than one\cite{Samukhin97}. The static conductivity on
such a nearly 1D fractal is also calculated to follow Eq.~(\ref{mott}) with
$\gamma=1/2$.

{\sl Model --} To include a transition from quasi-1D hopping to 3D hopping as
a function of doping level, the models mentioned above need to be extended. In
the quasi-1D model such a transition is expected when the transverse overlap
is sufficiently increased. In the nearly 1D fractal model, an increase in the
fractal dimension would eventually lead to a decrease of $\gamma$ down to
$1/4$. So in both cases, a transition to 3D hopping is induced by an increase
of the inter-chain connectivity. Before we present our calculations, let us
show how this can be qualitatively understood.  Upon chemically doping a
conjugated polymer, not only charge carriers, but also dopant counter-ions are
added to the system. The counter-ions locally decrease the inter-chain
potential seen by the carriers, and thus considerably enhances the hopping
rate\cite{Zuppiroli94}.  Since the Fermi energy is shifted upwards upon
doping, the counter-ions must even be considered as hopping sites when the
distance of the dopant site energy to the Fermi level becomes of the order of
the hopping energy. In this view, the sharply defined crossover at $c^{}_0$
can be interpreted as the point where the dopant sites start to play an active
role in the hopping process, thereby enhancing the density of hopping sites
and making variable range hopping in the direction perpendicular to the chains
possible.

We now show quantitatively that the conductivity in both doping regimes and
the crossover with doping level can be explained in a single variable-range
hopping model.  We assume hops within an energy interval $E$ over a distance
$(X,Z)$ in the combined parallel ($x$) and orthogonal ($z$) directions, with
localization lengths $L_x$ and $L_z$ respectively. For an electron on a chain,
the local DOS $n$ contains two contributions, $n^{}_0$ from the
chain itself and $n^{}_1$ from the neighboring chains and from intermediate
dopant sites. As the latter contribution depends stronger on $c$ than the
former, $n_1/n_0$ rises with doping level. Following the usual variable
range hopping arguments, the conductivity $\sigma \propto \exp[-2X/L_x -
2Z/L_z -E/k^{}_BT]$ should be maximized under the condition $2X (2Z)^2 \langle
n \rangle E \approx 1$, where $\langle n \rangle$ is the DOS averaged over the
volume $(2X,2Z,2Z)$, for which we write $\langle n \rangle = n^{}_0 (L_z/2Z)^2
+ n^{}_1$\cite{note1}.  We introduce $\xi = 2X/L_x$, $\zeta = 2Z/L_z$ and
$\epsilon = E/k^{}_B T$ and note that $\langle n \rangle$ does not depend on
$X$, leading to $\epsilon = \xi$, i.e.  $\sigma \propto \exp[-2\xi -\zeta]$.
Optimizing in the `high doping' limit $\zeta^2 \gg n^{}_0/n^{}_1$ we find
$2\xi = \zeta$ and $\xi^{-4} = 4 k^{}_B T L^{}_x L_z^2 n^{}_1$, which gives
$\sigma \propto \exp[-4\xi] = \exp[-(T_0^{high}/T)^{1/4}]$ with $T_0^{high} =
64/(k^{}_B L^{}_x L_z^2 n^{}_1)$.
For the `low-doping' regime $\zeta^2 \ll n^{}_0/n^{}_1$, we get
$\xi^{-2} = k^{}_B T L^{}_xL_z^2n^{}_0$ and $\zeta = 0$, which leads
to $\sigma \propto \exp[-2\xi] = \exp[-(T_0^{low}/T)^{1/2}]$ with $T_0^{low} =
4/(k^{}_B L^{}_x L_z^2 n^{}_0)$.
Note that although in this limit the $T$-dependence ($\gamma = 1/2$) is
determined by the dominating hops in the chain direction, occasional hops
between chains will still happen. In the high doping regime, hops in all
directions are equally likely and become long-ranged at low $T$.

{\sl AC conductivity --}
In Fig.~\ref{ACconductivity}, $\sigma(\omega)$ at room temperature is plotted
for three samples with doping levels $0.03 < c < 0.22$. At low frequencies,
the conductivity is seen to be independent of frequency and equal to the DC
value. At the onset frequency $\omega^{}_0/2\pi \approx 1$~MHz, the
conductivity starts to rise, following an approximate powerlaw $\sigma \sim
\omega^s$ with $s<1$. An extra upturn in the conductivity is observed around a
second frequency $\omega^{}_1/2\pi \approx 10$~ GHz.
The temperature dependence of the high frequency (200--600~GHz) conductivity
was also measured between 4 and 300~K (not shown), revealing that the
frequency dependence $\sigma  \propto \omega^s$ with $s \approx 1.6$ is
independent of temperature. The absolute value of the conductivity only shows
a weak (30\%) decrease going from 300 to 150~K, and is constant with
temperature below 150~K.

As was discussed above, the conductivity of a system of coupled polymer chains
generally consists of contributions due to both inter- and intra-chain
transport. When the chains are only weakly coupled, i.e. the inter-chain
hopping rate $\Gamma_{inter}$ is low, the conductivity $\sigma(\omega)$ at
frequencies $\omega \gg \Gamma_{inter}$ is dominated by charge transport
processes within single polymer chains. It is widely known that in a 1D chain,
any impurity causes the states to be localized, so the chain has zero
conductivity in the limit $\omega \rightarrow 0$, $T\rightarrow
0$\cite{Mott61}. At $T\rightarrow 0$ and $\omega>0$,
the conductivity of the chain is finite, stemming from resonant photon induced
transitions between the localized states, and is at low frequencies
$\omega\tau \ll 1$ given by\cite{Berezinskii74}

\begin{equation}
\sigma^{}_0(\omega) = {4\over{\pi\hbar b^2}}e^2 v^{}_F \tau (\omega\tau)^2
\ln^2(1/\omega\tau)
\label{1dmetal}
\end{equation}

\noindent{where $v^{}_F$ is the Fermi velocity on the chain, $\tau$ is the
backward scattering time and $b$ is the inter-chain separation. Following
Ref.\cite{Shklovskii81}, this may be written as $\sigma^{}_0(\omega) =
(\pi/2 b^2) e^2 g_0^2 L_x^3 \hbar\omega^2 \ln^2(1/\omega \tau )$,
where $g^{}_0 = n^{}_0 L_z^2$ is the on-chain DOS per unit length.  At
finite temperatures, an extra contribution is present due to phonon assisted
hopping within the chain. The phonon assisted conductivity is given by the
1D pair approximation\cite{Wang90}}

\begin{equation}
\sigma^{}_1(\omega,T) = {\pi^3\over {128 b^2}}e^2 g_0^2 L_x^3 k^{}_B T
\omega \ln^2(\nu_{ph}/\omega)
\label{pairappr}
\end{equation}

\noindent{valid for frequencies $\omega$ below the phonon `attempt' frequency
$\nu_{ph}$. The total conductivity at temperature $T$ is now given by the sum
of the two contributions, $\sigma(\omega,T) = \sigma^{}_0(\omega) +
\sigma^{}_1(\omega,T)$ given by Eqs.~(\ref{1dmetal}) and (\ref{pairappr}). The
data at $\omega/2\pi > 10$~MHz have been fitted with this $\sigma(\omega)$,
as is shown in the inset of Fig.~\ref{ACconductivity}. Here, the dielectric
loss function $\epsilon''(\omega) = \sigma(\omega)/\epsilon^{}_0\omega$ is
plotted at frequencies between 10~MHz and 3~THz, together with the fitting
line. The fits are excellent at high frequencies $\omega/2\pi > 200$~MHz.
The deviations below 200 MHz indicate that either multiple intra-chain
hopping or inter-chain transitions have significant contributions in the
MHz regime, which is consistent with the variable range hopping description
at low frequencies.

{\sl Parameter values --}
From the fits of the AC data, the parameters $\tau = 10^{-14}$~s and $\nu_{ph}
= 2\times 10^{12}$~s$^{-1}$ are extracted. While the phonon frequency
$\nu_{ph}$ is in good agreement with commonly suggested values of
$10^{12}$~s$^{-1}$\cite{Bottger85}, the scattering time $\tau$ is an
order of magnitude longer than reported values for other (highly conducting)
conjugated polymers\cite{Kohlman97,Lee95}, which is likely due to a smaller
$v^{}_F$ resulting from the low band filling. The typical time scales
now follow directly from the fits, since phonon mediated hops between two
sites within a chain occur at rates up to $\Gamma_{ph} = \nu_{ph}\exp[-
2X/L_x]$.  With $X/L_x \geq 1$, we have $\Gamma_{ph,max} \sim 10^{11}$~s$^{-
1}$, equivalent to a local diffusion constant $D = 10^{-7}$~m$^2$/s.

Assuming the localization lengths $L_x$ along the chain and
$L_z$ perpendicular to the chain to be independent of doping level, they
can be determined from the combined DC and high frequency conductivity data.
From the samples in the low-doping regime, $L_x$ and $g^{}_0$ can be extracted
using $T_0^{low} = 4/(k^{}_B L_x g^{}_0) \approx 10^5~$K and
Eq.~(\ref{pairappr}). This gives $L_x = 10$~\AA, indicating that carriers are
localized on the chains in regions consisting of two to three rings;
furthermore $g^{}_0 = 0.1$~levels/(eV ring) for $c=0.03$. For the typical
hopping distance along the chain we find $X = (L_x/4)(T_0^{low}/T)^{1/2} =
50$~\AA.  At the onset of the high-doping regime $\zeta^2 \approx 2
n^{}_0/n^{}_1$; using $Z \approx b$ and $T_0^{high} = 64/(k^{}_B L^{}_x L_z^2
n^{}_1) \approx 10^8$~K, we find $L_z = 1.4$~\AA, close to reported values for
other conjugated polymers lying deeply in the insulating
regime\cite{Li93,Vissenberg98}, and $n^{}_1 = 2\times 10^{26}$~eV$^{-1}$m$^{-
3}$, implying a DOS of 0.6 states per eV per dopant molecule. Within the
localized regions, the carriers move in the chain direction with velocity
$v_x$, which can be estimated using the fact that in 1D conductors
the mean free path $l_x = v_x \tau \sim L_x$; this gives $v_x
\sim 10^5$~m/s, similar to values observed in 1D-organic
conductors\cite{Soda77} and other conjugated polymers\cite{Beau99}.

In summary, we measured $\sigma(\omega)$ in a conjugated polythiophene with
small inter-chain overlap. We developed a model that allows a consistent
analysis of the $\sigma(\omega)$ data in terms of  inter- and intra-chain
transport.  From the low frequency results we have found that carriers are
strongly localized on 1D chains with $L_z = 1.0$~\AA, and no 3D metallic
islands are present. The high frequency data show 1D transport along the
polymer chains with a scattering time $\tau=10^{-14}$~s, while
intra-chain phonon assisted hopping proceeds at rates $\Gamma_{ph} \leq
10^{11}$~s$^{-1}$.

It is a pleasure to acknowledge B.F.M. de Waal who prepared the undoped
polythiophene samples, G.A. van Albada who assisted in the far-infrared
experiments, A. Goossens who performed the M\"ossbauer measurements, and L.J.
de Jongh and O. Hilt who were involved in the discussions. This research is
sponsored by the Stichting Fundamenteel Onderzoek der Materie, which is a part
of the Dutch Science Organization.


\begin{figure}[htb]
\begin{center}
\leavevmode
\epsfig{figure=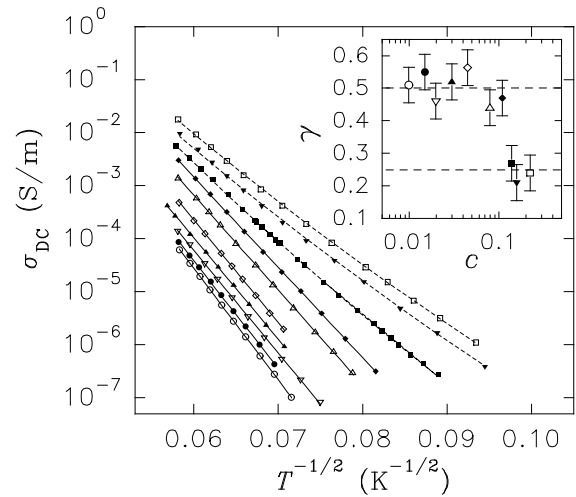,width=7.5cm,angle=0}
\end{center}
\caption{\noindent{DC conductivity as a function of temperature for several
doping levels. The data are fitted to  $\sigma^{}_{DC} = \sigma^{}_0 \exp[{-
(T^{}_0/T)^{\gamma}}]$. For the drawn lines $\gamma = 1/2$, whereas for the
dashed lines $\gamma=1/4$. The inset shows the $c$-dependence of $\gamma$ and
defines the symbols used in this graph.}
\label{DCconductivity}}
\end{figure}

\begin{figure}[htb]
\begin{center}
\leavevmode
\epsfig{figure=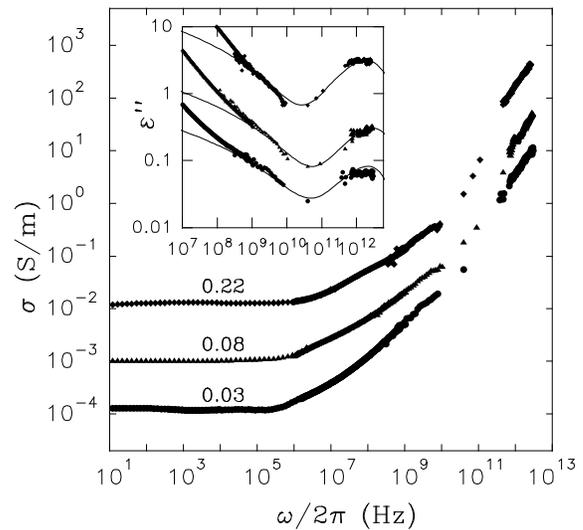,width=7.5cm,angle=0}
\end{center}
\caption{\noindent{$\sigma'(\omega)$ between 5~Hz and 3~THz at 300~K for
three doping levels $0.03<c<0.22$. In the inset the loss function
$\epsilon''(\omega) = \sigma'(\omega)/\epsilon^{}_0\omega$ is plotted
for the same three samples. Drawn lines are fits to Eqs.~(\ref{1dmetal}) and
(\ref{pairappr}). Data are corrected for the relatively small dipolar response
of the alkoxy side chain (the maximal value of $\epsilon''=0.1$ is centered
around 0.5 GHz), most clearly visible in the undoped polymer.} \label{ACconductivity}}
\end{figure}
}

\end{multicols}


\begin{thebibliography}{99}
\bibitem{film} Films prepared by drop casting had comparable $\sigma_{DC}$
values. Powders were more convenient for the experiments performed here.
\bibitem{note1} The number of states on a chain is given by $n_0\times
2XL_z^2$.  Therefore the single chain contribution to the density in
$(2X,2Z,2Z)$ drops by $(L_z/2Z)^2$.
\end{thebibliography}
\end{document}